\documentclass[reprint,amsmath,amssymb,aps,prl,showpacs]{revtex4-1}
\usepackage{amsmath}
\usepackage{graphicx}

\def\eqref#1{(\ref{#1})}

\def\be{\begin{equation}}
\def\ee{\end{equation}}
\def\bea{\begin{eqnarray}}

\def\eea{\end{eqnarray}}
\def\nn{\nonumber}

\def\le{\left}
\def\ri{\right}

\def\6{\partial}

\def\re{\mathrm{Re}}

\def\Sq{S}
\def\Sqt{\widetilde S}
\def\flf{\Xi}

\begin{document}

\title{Quasi-normal modes of a strongly coupled non-conformal plasma and approach to criticality}

\author{Panagiotis Betzios}
\email{P.Betzios@uu.nl}
\affiliation{Institute for Theoretical Physics and Center for Extreme Matter and Emergent Phenomena, Utrecht University, Leuvenlaan 4, 3584 CE Utrecht, The Netherlands}
\author{Umut G\"ursoy}
\email{U.Gursoy@uu.nl}
\affiliation{Institute for Theoretical Physics and Center for Extreme Matter and Emergent Phenomena, Utrecht University, Leuvenlaan 4, 3584 CE Utrecht, The Netherlands}
\author{Matti J\"arvinen}
\email{jarvinen@lpt.ens.fr}
\affiliation{Institut de Physique Th\'eorique Philippe Meyer \& Laboratoire de Physique Th\'eorique, 
\'Ecole Normale Sup\'erieure, PSL Research University, CNRS, 24 rue Lhomond, 75005 Paris, France}
\author{Giuseppe Policastro}
\email{policast@lpt.ens.fr}
\affiliation{Laboratoire de physique th\'eorique, D\'epartement de physique de  l' ENS, \'Ecole normale sup\'erieure PSL Research University, Sorbonne Universités, UPMC Univ. Paris 06, CNRS, 75005 Paris, France}
\date{\today}

\begin{abstract} We study fluctuations around equilibrium in a class of strongly interacting non-conformal plasmas using holographic techniques. In particular we calculate the quasi-normal mode spectrum of black hole backgrounds that approach to Chamblin-Reall plasmas in the IR. In a specific limit, related to the exact linear-dilaton background in string theory, we observe that the plasma approaches criticality and we obtain the quasi-normal spectrum analytically. 
We regulate the critical limit by gluing the IR geometry that corresponds to the non-conformal plasma to a part of AdS space-time in the UV.
Near criticality, we find two sets of quasi-normal modes, related to the IR and UV parts of the geometry. In the critical limit, the quasi-normal modes accumulate to form a branch cut in the correlators of the energy-momentum tensor on the real axis of the complex frequency plane.
\end{abstract}

\maketitle

\section{Introduction}

Many applications of holographic duality concern systems around a critical point, where there is scale invariance. However many realistic systems are not scale invariant in the regime of interest, QCD being a prominent example. It is then very interesting to study how the absence of scale invariance influences the transport properties of such systems in a controlled manner. 

A simple class of models that allows this investigation is Einstein gravity coupled to a dilaton with an exponential potential $\propto e^{\alpha \phi}$, for which analytical black hole solutions are known \cite{ChamblinReall}. We call the corresponding finite temperature system the \emph{Chamblin-Reall (CR) plasma}. The parameter $\alpha$ determines the beta function of the dual theory and the deviation from conformality. At $\alpha=0$ we recover the conformal theory. This form of the potential is a good approximation for the IR behavior of a class of QCD-like holographic models \cite{ihqcd1,ihqcd2}. 

In a previous paper \cite{Gursoy:2015nza} we studied time-dependent solutions of this model corresponding to boost-invariant flow, and found that the system has a slower approach to equilibrium than in the 
conformal case. In this letter we continue this investigation, focusing on the quasi-normal modes that contain information about the thermalization processes that lead to equilibrium. We are interested in the behavior of the modes  as a function of the parameter $\alpha$. For $\alpha=0$ the spectrum is well-known \cite{Starinets:2002br}: the modes lie along straight lines in the lower complex frequency half-plane.   

We notice that there is a critical value $\alpha_c = 4/3$ for which the string frame geometry matches with 
the 2D dilatonic black hole geometry, the linear dilaton background  (plus some extra flat coordinates). The change in the geometry signals a phase transition, that can be shown to be continuous, of the BKT type \cite{Gursoy:2010jh} \footnote{In order to make the statement precise, one has to modify  the potential slightly.}. 

We have computed the spin-2 quasi-normal  spectrum numerically, and analytically in the vicinity of the phase transition. 
We will present the main findings in this letter, concentrating in particular on the analytic results in the limit $\alpha \to \alpha_c$. The details of the derivation and a more extensive presentation of the results will be given elsewhere \cite{WIP}. In summary, we see that when we approach the critical point, the quasi-normal modes become closely spaced and their imaginary part decreases, until eventually they form a branch cut on the real axis, even as the temperature remains finite at the critical point. We argue that this behavior signals a breakdown of hydrodynamics, as there are modes that decay more slowly than the hydrodynamic fluctuations. However it is not clear if there is a dual theory exactly at the critical point (it has been suggested that it could correspond to an order-disorder transition in a spin model \cite{Gursoy:2010kw}). 

We can have a well-defined dual theory if we consider a  completion obtained by replacing part of the UV geometry with a slice of AdS spacetime. The gluing introduces a new scale, and even at the critical point the modes are discrete; we observe that there are two distinct sets of modes with different behavior.  There is a set which includes the lowest-lying modes, that are related to those of the CR geometry; these modes approach the CR modes as the temperature is lowered (or equivalently the scale of the UV completion is increased). There is also another set of modes, starting at a higher frequency whose imaginary part is only mildly frequency-dependent. We do not have a good understanding of the presence of these modes, but we can conclude that the CR modes give a good approximation to the lowest-lying spectrum of the completed theory.

\section{Quasi-normal modes of the CR plasma}
 
We consider five dimensional Einstein-dilaton gravity
\begin{equation}\label{action}
 {\cal A} = \frac{1}{16\pi G_5} \int d^{5}x \sqrt{-g}\left( R - \frac43 (\6\phi)^2 
+V_0 e^{\alpha \phi}
  \ri)\, .
\end{equation}
Defining
\be \label{fxidef}
 f(\hat r) = 1 - \left(\frac{\hat r}{\hat r_h}\right)^\xi \ , \qquad \xi = \frac{4(1-9 \alpha^2/64)}{1-9\alpha^2/16} \ ,
\ee
so that $\xi \to \infty$ at the critical point $\alpha_c =4/3$, 
the exact CR solution of the model is given in the Eddington-Finkelstein coordinates as
\bea
 ds^2 &=& e^{2A_0} \hat r^{-\frac{2}{3}{\xi-1}}\left[-2 \ell' d\hat r dv -f(\hat r) dv^2 + \delta_{ij} dx^{i} dx^{j}\right] \nonumber\\
 \phi &=& \frac{1}{2}\sqrt{(\xi-1)(\xi-4)} \, \log \hat r
\eea
where $\ell' = 3 \xi/\sqrt{V_0(1-1/\xi)}$. For this geometry the temperature reads
\be \label{Tdef}
  T  = \frac{\xi}{4\pi \hat r_h \ell'} \ . 
\ee

We now consider the spin-two fluctuations of the metric with momentum $k$ along the $x^1$-direction (the excitations being transverse to $k$). These can be shown to satisfy the relatively simple equation
\begin{eqnarray} \label{flucteq}
\hat r f(\hat r) \flf''(\hat r)+\left(2 i \ell' \hat r  \omega +f(\hat r)-\xi\right)\flf'(\hat r) && \nonumber \\
- \left(\ell'^2 k^2 \hat r + (\xi-1) \ell' i \omega \right) \flf(\hat r) &=& 0 \ .
\end{eqnarray}
All other fluctuations of the system also satisfy this same equation in the limit $k \to 0$.

Since the linear dilaton model is exactly solvable, we can find analytical results for the location quasi-normal modes in a perturbation expansion in $\xi^{-1} \sim \alpha - \alpha_c$, which are corroborated by the numerical results. 
In particular, we can compute explicitly the analytic result for the correlator of the transverse components of the energy-momentum tensor to leading order in $1/\xi$.
In order to write down our result for the correlator we define 
the reflection amplitude
\be \label{Rampl}
 \mathcal{R}(\varpi,q) = - \frac{\Gamma \left(1+i\Sqt\right)\, \Gamma \left(\frac{1}{2} \left(1-i \varpi -i\Sqt\right)\right)^2}{ \Gamma \left(1-i\Sqt\right)\, \Gamma \left(\frac{1}{2} \left(1-i \varpi +i\Sqt\right)\right)^2} \ ,
\ee
where 
\be
 \Sqt = \sqrt{\varpi^2-q^2-1}
\ee
with $\varpi = \omega/(2\pi T)$, $q=k/(2\pi T)$. The reflection amplitude arises from the behavior of the fluctuation in the deep IR region of the geometry and is similar to what is found in the minisuperspace studies of the 2D black hole of the linear dilaton model \cite{Dijkgraaf:1991ba,Nakayama:2007sb}. 
The final result for the correlator is however more complicated due to a nontrivial interplay of the fluctuations in the UV and in the IR. In terms of the reflection amplitude the leading order result reads \cite{WIP}
\bea \label{corrfinal}
 G_\mathrm{reg} &=& \frac{2\pi\, \xi^\xi \hat r_h^{-\xi} }{\Gamma\left(\frac{\xi}{2}\right)\Gamma\left(1+\frac{\xi}{2}\right)}\left(\frac{\left(\varpi ^2-q^2\right)}{16}\right)^\frac{\xi}{2} \nn \\
 &\times& \left[i+\left(\frac{1+i \Sqt}{1-i \Sqt}\right)^\frac{\xi}{2} \frac{e^{-i \xi \Sqt}}{\mathcal{R}}\right]^{-1}
\eea
for $\re \varpi \gtrsim \sqrt{1+q^2}$, and
\bea \label{corrsmallomega}
  G_\mathrm{reg} &=& \frac{2\pi\, \xi^\xi \hat r_h^{-\xi}}{\Gamma\left(\frac{\xi}{2}\right)\Gamma\left(1+\frac{\xi}{2}\right)} \left(\frac{\left(\varpi ^2-q^2\right)}{16}\right)^\frac{\xi}{2} \nn \\
&\times&  \left[\left(\frac{1+\Sq}{1-\Sq}\right)^\frac{\xi}{2}\, e^{-\xi \Sq}\, \mathcal{R}- i\theta(-\mathrm{Im}\,\varpi) \right] 
\eea
for $\re \varpi \lesssim \sqrt{1+q^2}$. Here the subscript ``$\textrm{reg}$" indicates that we have subtracted a singular term which does not contain any information on the quasi-normal modes. We chose the branches of the square root factors such that the solution for negative $\varpi^2-q^2-1$ is given by replacing $\Sqt \mapsto  i\Sq$ where
\be
 \Sq = \sqrt{q^2-\varpi^2+1}\ ,
\ee
which corresponds to analytic continuation through the upper half of the complex $\varpi$-plane.

\begin{figure}[!tb]
\begin{center}
\includegraphics[width=0.45\textwidth]{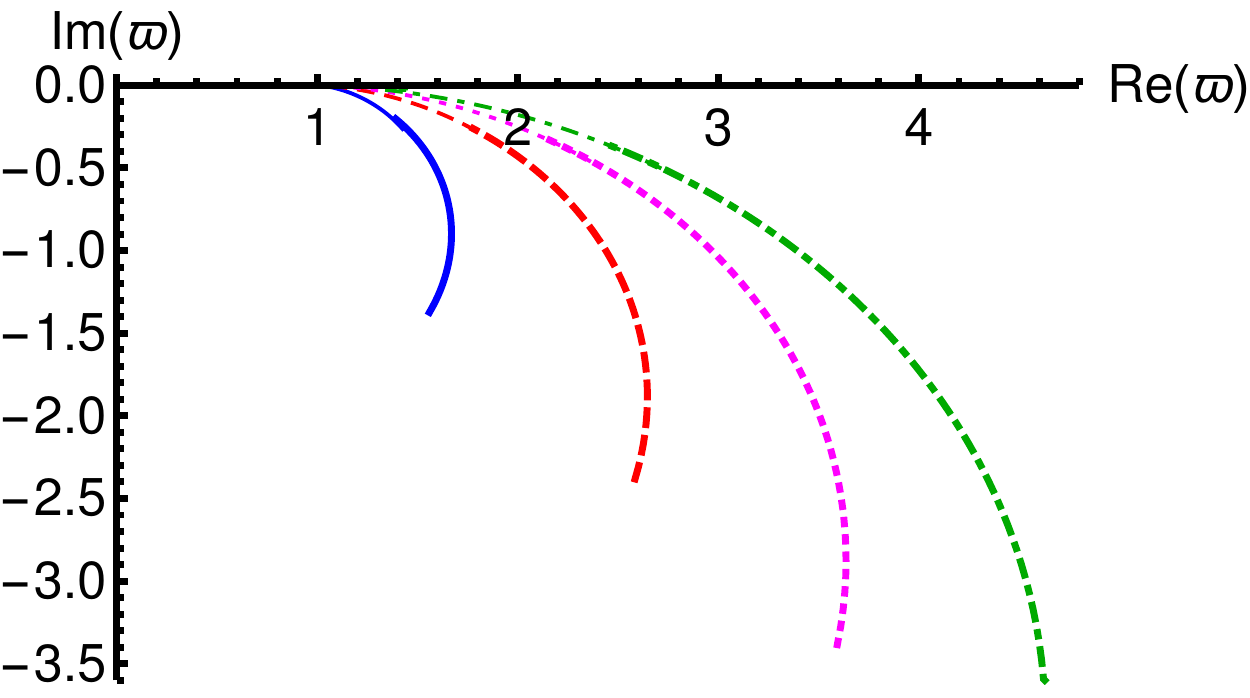}%
\end{center}
\caption{The dependence of the four lowest nonhydrodynamic quasi-normal modes on $\xi$ at $q=0$. Thick lines were obtained by directly solving the fluctuation equation~\eqref{flucteq} numerically for $4\leq \xi \lesssim 20$, and the thin lines are based on the analytic approximation in~\eqref{xiinftyG}. }
\label{fig:modexdep}
\end{figure}

The former expression~\eqref{corrfinal} has a series of poles corresponding to the nonhydrodynamic quasi-normal modes of the system.
The main result is that the quasi-normal modes become closely spaced and approach the real axis as $\xi \to \infty$. In the limit, they should create a branch cut on the real axis running from $\varpi =\sqrt{1+q^2}$ to $\varpi = \infty$. This can be seen for the four lowest modes in Fig. \ref{fig:modexdep}. 

The hydrodynamic quasi-normal modes by contrast do not scale with $\xi$: the imaginary part is fixed by the shear viscosity/entropy ratio that is $1/4\pi$ independently of $\xi$. This means that for arbitrarily small momenta one can find non-hydro modes that are decaying slower than the hydro ones; in other words, hydrodynamics breaks down on arbitrarily large scales in the limit $\xi \to \infty$.

\section{UV completion}

However the critical limit is subtle. It is not possible to send $\xi \to \infty$ in eq. (\ref{corrfinal}) because of wildly oscillating phases in the denominator. Thus at the critical point we do not have a well-defined prescription for computing correlators. This can be understood as a consequence of  the fact that the geometry becomes asymptotically flat in the presence of a linear dilaton; the asymptotic behavior of the fluctuations allows instead  to define a scattering S-matrix corresponding to the reflection amplitude (\ref{Rampl}).  The definition of the boundary theory requires a UV regulator and the geometry being asymptotically AdS. For this reason, and also because the CR plasma is anyway only supposed to be an accurate description in the IR, we turn to the question of the robustness of our result when the theory is embedded into a UV complete description. 

\begin{figure}[!tb]
\begin{center}
\includegraphics[width=0.48\textwidth]{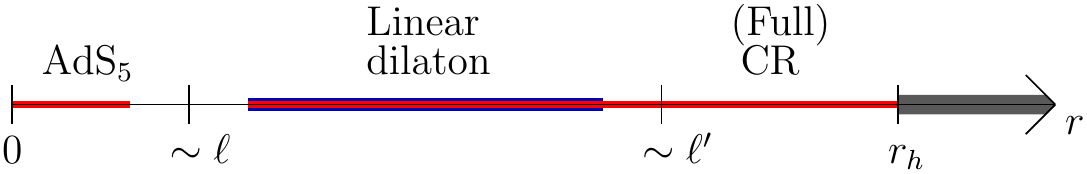}%
\end{center}
\caption{The structure of the 
geometry for potentials asymptoting to CR behavior in the IR at small temperatures with large $\xi$. }
\label{fig:geometry}
\end{figure}

In order to implement this completion, we should consider dilaton potentials which admit a flow from AdS geometry in the UV to the CR geometry in the IR.
At large $\xi$ and small enough $T$ the geometry has the structure depicted in Fig. \ref{fig:geometry}. The  radial coordinate $r$ is related to $\ell'\hat r$ of the previous section by a shift that is needed to set the boundary at $r=0$. There are three regions: an AdS in the UV ($0<r\ll \ell$), the CR geometry in the IR ($r \gg \ell$), and an intermediate region where the linear dilaton approximation is valid ($\ell \ll r \ll \ell'$ with $\ell' \sim \ell\xi$); the extent of the middle region grows with $\xi$, so we expect that at least some of the QNM of the CR geometry will approximate those of the full geometry when $\xi$ is large. However the quasinormal modes now acquire a non-trivial dependence on the temperature as well as on $\xi$.

\begin{figure*}[!tb]
\begin{center}
\includegraphics[width=0.45\textwidth]{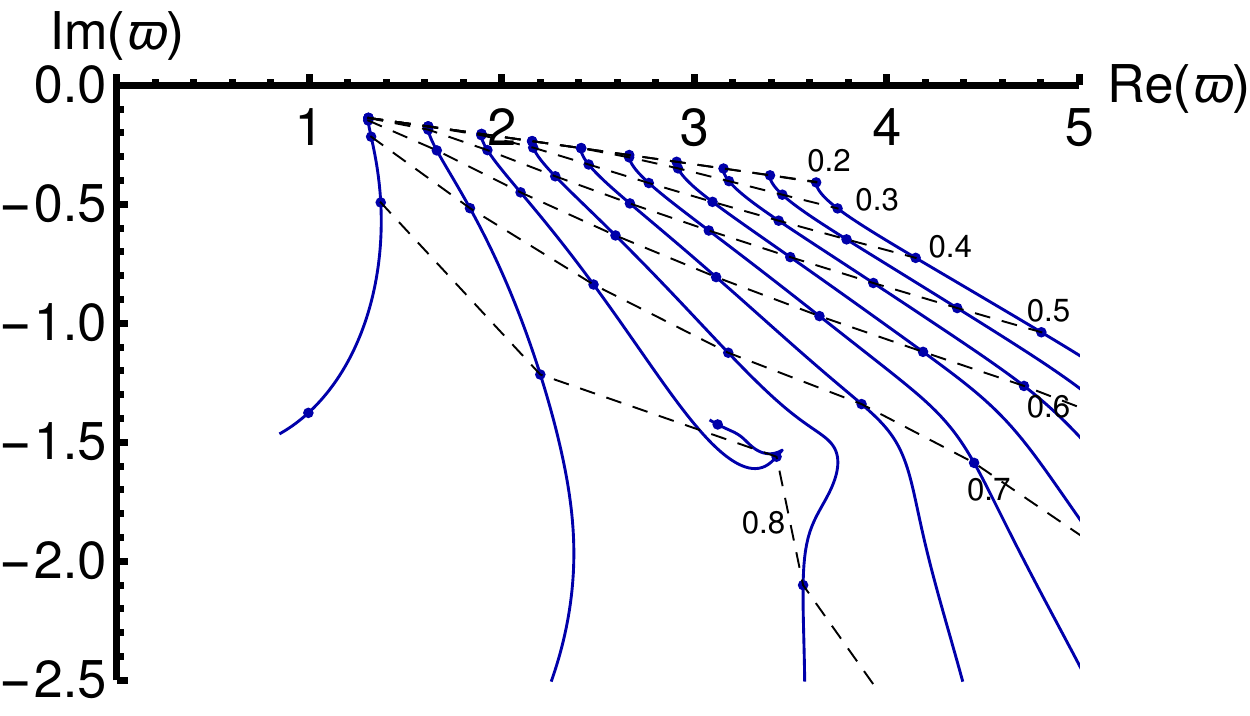}
\hspace{4mm}\includegraphics[width=0.45\textwidth]{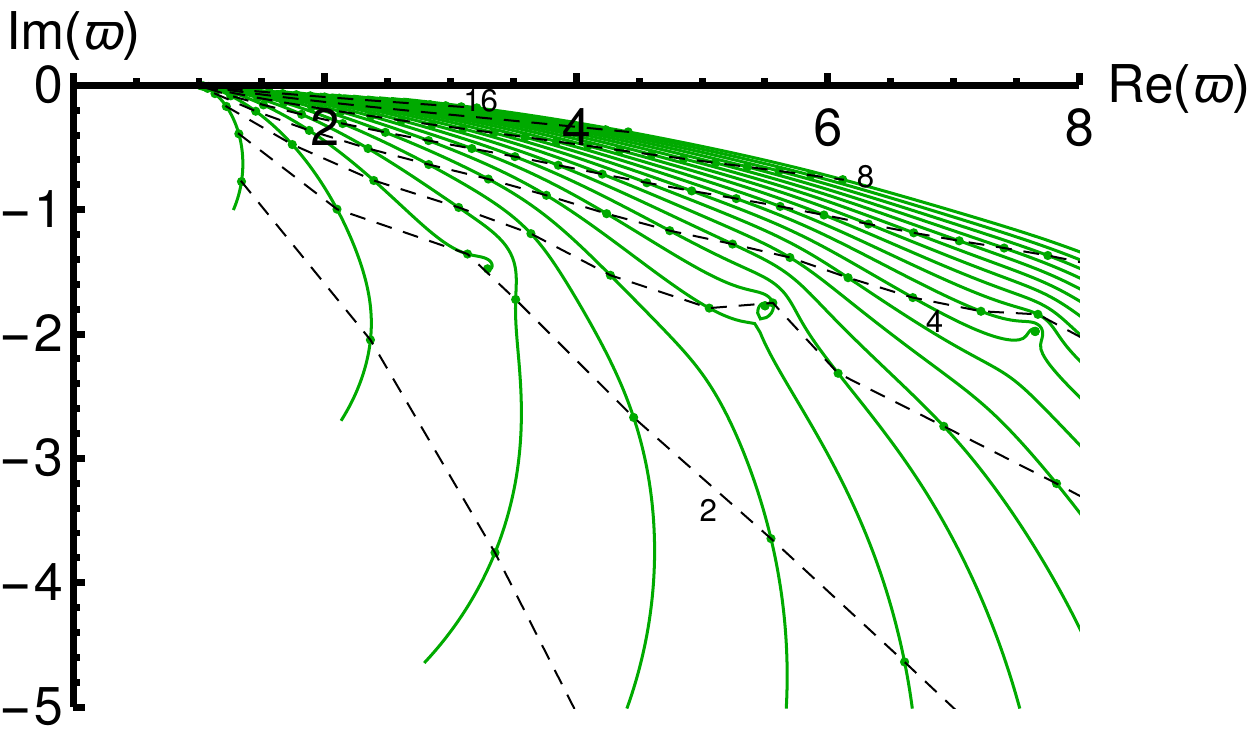}
\end{center}
\caption{The dependence of the location of quasi-normal modes on the location of the horizon at $q=0$ in the setup where AdS and CR geometries were glued together. Left: The trajectories of the ten lowest QNMs on the complex $\varpi$-plane  at $\xi\simeq 26.77$ as $T$ grows from $T=0.2 T_c$ to $T=0.91 T_c$. The dashed curves were added to guide the eye, they are at constant $T/T_c$ with values of the ratio indicated by the labels. The markers are at $T/T_c=0.2$, $0.3$, \ldots $0.9$ for all curves. Right: The trajectories of the QNMs at $\xi =\infty$ on the complex $\varpi$-plane as $r_h$ is varied from $r_h = 1.2 r_c$ (lower end points of the curves) to $r_h=20 r_c$ (upper end points). The dots are the locations of the QNMs for $r_h/r_c = \sqrt{2}$, $2$, $2\sqrt{2}$, \ldots $16$. The dashed lines  connect the locations at the same $r_h$ as indicated by the labels.}
\label{fig:QNMevolution}
\end{figure*}

For simplicity, instead of changing the potential, we adopt a prescription that consists in gluing the CR solution to an AdS space in the UV, 
 so that the transition region around $r \sim \ell$ in Fig. \ref{fig:geometry} is replaced by a junction at some arbitrary point $r=r_c$.
 In this approximation and to leading order in $1/\xi$ the spin-two correlator can again be found analytically (but the expression is too complicated to reproduce it in this letter).
The modes of the completed geometry extracted from the analytic result are shown in Fig. \ref{fig:QNMevolution} (left). They
indeed depend on the temperature and approach the modes of the CR geometry only at low temperature. 
The cusps are related to appearance of the second set of modes at higher temperatures which we will discuss in more detail below.
As the temperature grows the horizon $r_h$ moves to the AdS region, and the 
junction approximation breaks down. This happens at $T=T_c$ where $T_c = \xi/4\pi\ell'$ is the temperature of the linear dilaton solution.

\begin{figure*}[!tb]
\begin{center}
\includegraphics[width=0.45\textwidth,trim=0 11mm 0 0]{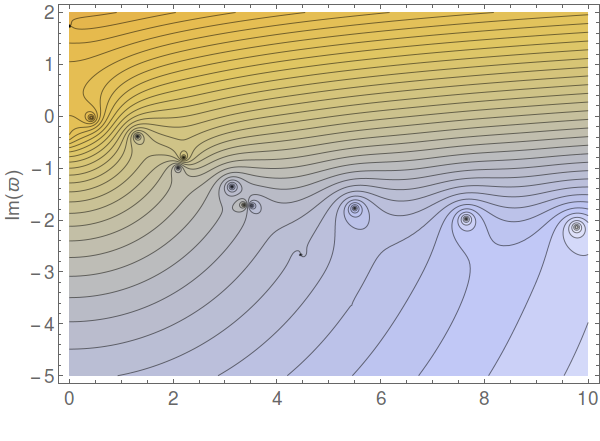}%
\hspace{4mm}\includegraphics[width=0.43\textwidth]{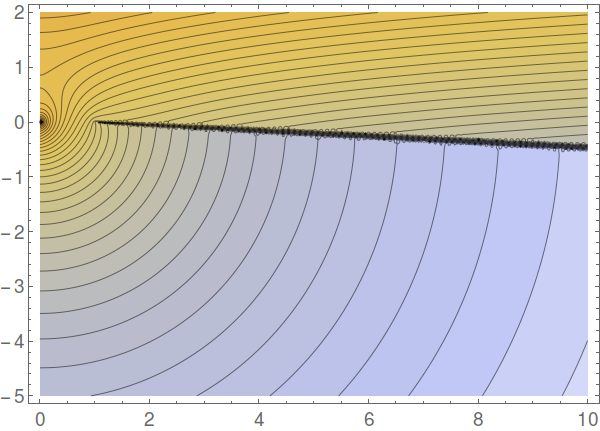}

\vspace{2mm}

\includegraphics[width=0.45\textwidth]{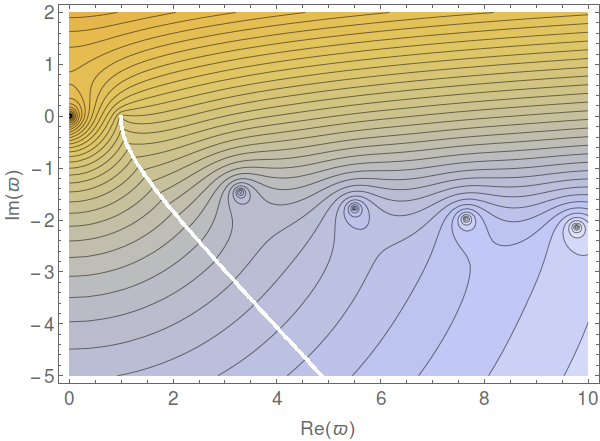}%
\hspace{4mm}\includegraphics[width=0.43\textwidth]{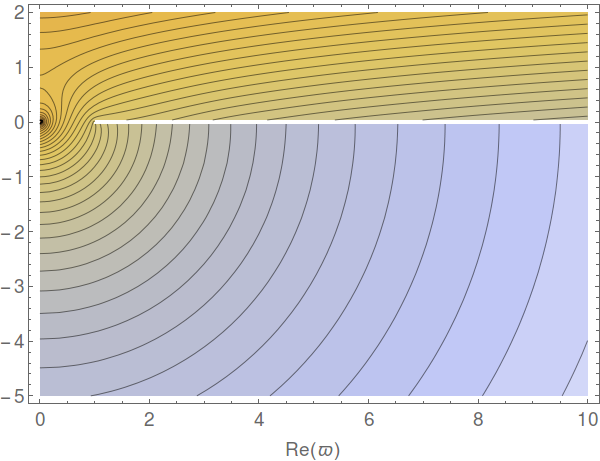}
\end{center}
\caption{The (logarithm of the) absolute value of the correlators of the energy-momentum tensor on the complex $\varpi$-plane in various approximations. The plots are for $\xi=\infty$ and $q=0$. The contours are at constant values of $|\widetilde G_\mathrm{reg}|$, with orange/yellow colors (mostly top parts of the plots) indicating small values and blue/white colors  (mostly bottom parts of the plots) indicating large values. Top left: the ``glued'' correlator of~\protect\eqref{xiinftyG} at $r_h/r_c = 2$. Top right: the correlator of~\protect\eqref{xiinftyG} at $r_h/r_c = 20$. Bottom left: the large $\varpi$ approximation of the correlator with $r_h/r_c = 2$ \cite{WIP}. Bottom right: the limit of large black hole~\protect\eqref{rhinftylimit} of the glued correlator. }
\label{fig:gluedcorrelator}
\end{figure*}

\section{Approach to criticality}

In the UV-completed geometry we can take the critical limit $\xi \to \infty$ as there is no issue with the boundary conditions. 
In this limit, the transverse spin-two correlator simplifies to
\begin{widetext}
\begin{align} \label{xiinftyG}
\widetilde G_\mathrm{reg} = -\frac{81 i \pi  \hat \mu^4}{512 r_c^4 }  \frac{\hat \mu  \left[1+e^{3 \Sq\le(1-\frac{r_h}{r_c}\ri)} \mathcal{R}\right]H_1^{(1)}\left(\frac{3 \hat \mu}{2}\right)+ \left[(\Sq-1) -e^{3 \Sq\le(1-\frac{r_h}{r_c}\ri)} (\Sq+1) \mathcal{R}\right]H_2^{(1)}\left(\frac{3 \hat \mu}{2}\right)}{\hat \mu  \left[1+e^{3 \Sq\le(1-\frac{r_h}{r_c}\ri)} \mathcal{R}\right]J_1\left(\frac{3 \hat \mu}{2}\right)+ \left[(\Sq-1)-e^{3 \Sq\le(1-\frac{r_h}{r_c}\ri)} (\Sq+1) \mathcal{R}\right]J_2\left(\frac{3 \hat \mu}{2}\right)} \ ,
\end{align}
\end{widetext}
where $\hat \mu 
=\sqrt{\varpi^2 -q^2}$, $J_i$ are Bessel functions, and $H^{(1)}_i$ are Hankel functions of the first kind. 
The branch cut arising from the square root in the definition of $\Sq$ cancels in this expression: it is invariant under $\Sq \mapsto -\Sq$ (which implies $\mathcal{R} \mapsto \mathcal{R}^{-1}$). Therefore the only singularities are poles due to the quasi-normal modes.

There is a subtlety regarding the temperature in the critical limit. Namely, at any finite $\xi$ the temperature is a monotonic function of $r_h$, but in the critical limit the dependence of the temperature on the location of the horizon disappears and it is fixed to $T_c$ \cite{Gursoy:2015nza}. Our gluing procedure works when $r_h \gg r_c$.

The structure of the correlator is depicted in Fig.  \ref{fig:gluedcorrelator}. We see that at small $r_h$ %high temperature 
there are two sets of poles: one set lying on an approximately straight line, and another one for which the imaginary part is almost frequency-independent. 
As $r_h$ is increased 
the slope of the first set of modes decreases and eventually reaches the real axis; for this reason we identify these poles as being associated with the CR part of the geometry, whereas the other poles are pushed to higher frequency and eventually disappear. The trajectories of the quasi-normal modes as $r_h$ varies are shown in  Fig. \ref{fig:QNMevolution} (right). As we pointed out above the cusps in the trajectories are related to the second set of modes: as $r_h$ decreases some of the CR modes stop their evolution when $\mathrm{Re}\, \varpi \sim r_h/r_c$ and join the line of the second set of modes.

The appearance of the branch cut in the limit of large black hole can also be seen analytically by taking the limit $r_h \to \infty$ of~\eqref{xiinftyG}. We obtain
\be \label{rhinftylimit}
 \widetilde G_\mathrm{reg} \to -\frac{81 i \pi  \hat \mu^4}{512 r_c^4 }  \frac{\hat \mu H_1^{(1)}\left(\frac{3 \hat \mu}{2}\right)+ (\Sq-1) H_2^{(1)}\left(\frac{3 \hat \mu}{2}\right)}{\hat \mu J_1\left(\frac{3 \hat \mu}{2}\right)+ (\Sq-1) J_2\left(\frac{3 \hat \mu}{2}\right)} \ .
\ee
This expression is no longer invariant under $\Sq \mapsto -\Sq$ and therefore we find a physical branch cut running from $\varpi = \sqrt{1+q^2}$ to $\varpi = \infty$. This branch cut is visible in the plots on Fig. \ref{fig:gluedcorrelator} (right). 

\section{Conclusions}

Our results should be compared to \cite{Buchel:2015saa, Attems:2016ugt, Janik:2015waa, Ishii:2015gia} who also studied the dependence of the imaginary part of the lowest QNM in systems with broken scale invariance, but found a milder dependence on the scale-breaking parameter.  

It would be very interesting to understand if it is possible to make sense of the critical limit independently of the UV completion, especially in light of possible condensed matter applications, as we mentioned in the introduction. In particular we wonder if there is a generic model for the breakdown of the hydrodynamic description that we observe in our system. 

Another interesting question is whether our result can have phenomenological applications to the QCD quark-gluon plasma. The role of the bulk viscosity, which should be important close to the deconfinement transition, has been discussed in this context (e.g. in \cite{Ryu:2015vwa}). A natural direction for further research would be to try and build a picture of how the long lived non-hydrodynamic modes affect the long-time behavior of the system. 

It is worth mentioning, even though it is not directly related to our work, that the structure of the quasi-normal spectrum that we found bears a resemblance to the results of \cite{Anninos:2017hhn} who attempted to make holographic sense of the quasi-normal modes of 2d de Sitter space by completing it in the UV with $AdS_2$. 

Even though we studied the fluctuations at non-zero momentum, in this letter we reported only on the modes at zero momentum. The behaviour of the quasi-normal modes as function of $q$ will be discussed in \cite{WIP}. 

\section{Acknowledgments} 

We would like to thank D. Anninos, R. Janik, E. Kiritsis, D. Mateos, and D.T. Son for discussions and helpful suggestions. This work is partially supported by the Netherlands Organisation for Scientific Research (NWO) under the VIDI grant 680-47-518, and the Delta-Institute for Theoretical Physics (D-ITP) that is funded by the Dutch Ministry of Education, Culture and Science (OCW).


\begin{thebibliography}{1}%
\makeatletter
\providecommand \@ifxundefined [1]{%
 \@ifx{#1\undefined}
}%
\providecommand \@ifnum [1]{%
 \ifnum #1\expandafter \@firstoftwo
 \else \expandafter \@secondoftwo
 \fi
}%
\providecommand \@ifx [1]{%
 \ifx #1\expandafter \@firstoftwo
 \else \expandafter \@secondoftwo
 \fi
}%
\providecommand \natexlab [1]{#1}%
\providecommand \enquote  [1]{``#1''}%
\providecommand \bibnamefont  [1]{#1}%
\providecommand \bibfnamefont [1]{#1}%
\providecommand \citenamefont [1]{#1}%
\providecommand \href@noop [0]{\@secondoftwo}%
\providecommand \href [0]{\begingroup \@sanitize@url \@href}%
\providecommand \@href[1]{\@@startlink{#1}\@@href}%
\providecommand \@@href[1]{\endgroup#1\@@endlink}%
\providecommand \@sanitize@url [0]{\catcode `\\12\catcode `\$12\catcode
  `\&12\catcode `\#12\catcode `\^12\catcode `\_12\catcode `\%12\relax}%
\providecommand \@@startlink[1]{}%
\providecommand \@@endlink[0]{}%
\providecommand \url  [0]{\begingroup\@sanitize@url \@url }%
\providecommand \@url [1]{\endgroup\@href {#1}{\urlprefix }}%
\providecommand \urlprefix  [0]{URL }%
\providecommand \Eprint [0]{\href }%
\providecommand \doibase [0]{http://dx.doi.org/}%
\providecommand \selectlanguage [0]{\@gobble}%
\providecommand \bibinfo  [0]{\@secondoftwo}%
\providecommand \bibfield  [0]{\@secondoftwo}%
\providecommand \translation [1]{[#1]}%
\providecommand \BibitemOpen [0]{}%
\providecommand \bibitemStop [0]{}%
\providecommand \bibitemNoStop [0]{.\EOS\space}%
\providecommand \EOS [0]{\spacefactor3000\relax}%
\providecommand \BibitemShut  [1]{\csname bibitem#1\endcsname}%
\let\auto@bib@innerbib\@empty
%</preamble>
\bibitem [{Note1()}]{Note1}%
  \BibitemOpen
  \bibinfo {note} {In order to make the statement precise, one has to modify
  the potential slightly.}\BibitemShut {Stop}%
\end{thebibliography}%


\begin{thebibliography}{99}

\bibitem{ChamblinReall}
  H.~A.~Chamblin and H.~S.~Reall,
  %``Dynamic dilatonic domain walls,''
  Nucl.\ Phys.\  B {\bf 562} (1999) 133
  [hep-th/9903225].
  %%CITATION = NUPHA,B562,133;%%
  
  
\bibitem{ihqcd1}
  U.~G\"ursoy and E.~Kiritsis,
  %``Exploring improved holographic theories for QCD: Part I,''
  JHEP {\bf 0802} (2008) 032
  [arXiv:0707.1324 [hep-th]].
  %%CITATION = ARXIV:0707.1324;%%
  %212 citations counted in INSPIRE as of 25 May 2015
  

\bibitem{ihqcd2}
  U.~G\"ursoy, E.~Kiritsis and F.~Nitti,
  %``Exploring improved holographic theories for QCD: Part II,''
  JHEP {\bf 0802} (2008) 019
  [arXiv:0707.1349 [hep-th]].
  %%CITATION = ARXIV:0707.1349;%%
  %223 citations counted in INSPIRE as of 25 May 2015


%\cite{Gursoy:2015nza}
\bibitem{Gursoy:2015nza} 
  U.~G\"ursoy, M.~J\"arvinen and G.~Policastro,
  %``Late time behavior of non-conformal plasmas,''
  JHEP {\bf 1601}, 134 (2016)
%  doi:10.1007/JHEP01(2016)134
  [arXiv:1507.08628 [hep-th]].
  %%CITATION = doi:10.1007/JHEP01(2016)134;%%
  %4 citations counted in INSPIRE as of 09 Apr 2016  

%\cite{Starinets:2002br}
\bibitem{Starinets:2002br}
  A.~O.~Starinets,
  %``Quasinormal modes of near extremal black branes,''
  Phys.\ Rev.\ D {\bf 66} (2002) 124013
 % doi:10.1103/PhysRevD.66.124013
  [hep-th/0207133].

%\cite{Gursoy:2010jh}
\bibitem{Gursoy:2010jh}
  U.~G\"ursoy,
  %``Continuous Hawking-Page transitions in Einstein-scalar gravity,''
  JHEP {\bf 1101} (2011) 086
 % doi:10.1007/JHEP01(2011)086
  [arXiv:1007.0500 [hep-th]].
  
 \bibitem{WIP}
P.~Betzios, U.~G\"ursoy, M.~J\"arvinen and G.~Policastro, {\it work in progress}


%\cite{Gursoy:2010kw}
\bibitem{Gursoy:2010kw}
  U.~G\"ursoy,
  %``Gravity/Spin-model correspondence and holographic superfluids,''
  JHEP {\bf 1012} (2010) 062
%  doi:10.1007/JHEP12(2010)062
  [arXiv:1007.4854 [hep-th]].

%\cite{Dijkgraaf:1991ba}
\bibitem{Dijkgraaf:1991ba}
  R.~Dijkgraaf, H.~L.~Verlinde and E.~P.~Verlinde,
  %``String propagation in a black hole geometry,''
  Nucl.\ Phys.\ B {\bf 371} (1992) 269.
%  doi:10.1016/0550-3213(92)90237-6


%\cite{Nakayama:2007sb}
\bibitem{Nakayama:2007sb}
  Y.~Nakayama,
 % ``Black Hole: String transition and rolling D-brane,''
  hep-th/0702221.  
  
  %\cite{Buchel:2015saa}
\bibitem{Buchel:2015saa}
  A.~Buchel, M.~P.~Heller and R.~C.~Myers,
  %``Equilibration rates in a strongly coupled nonconformal quark-gluon plasma,''
  Phys.\ Rev.\ Lett.\  {\bf 114} (2015) 25,  251601
  [arXiv:1503.07114 [hep-th]].
  %%CITATION = ARXIV:1503.07114;%%
  
  %\cite{Attems:2016ugt}
\bibitem{Attems:2016ugt}
  M.~Attems, J.~Casalderrey-Solana, D.~Mateos, I.~Papadimitriou, D.~Santos-Olivn, C.~F.~Sopuerta, M.~Triana and M.~Zilho,
  %``Thermodynamics, transport and relaxation in non-conformal theories,''
  JHEP {\bf 1610} (2016) 155
%  doi:10.1007/JHEP10(2016)155
  [arXiv:1603.01254 [hep-th]].


  
  %\cite{Janik:2015waa}
\bibitem{Janik:2015waa}
  R.~A.~Janik, G.~Plewa, H.~Soltanpanahi and M.~Spalinski,
  %``Linearized nonequilibrium dynamics in nonconformal plasma,''
  Phys.\ Rev.\ D {\bf 91} (2015) 12,  126013
  [arXiv:1503.07149 [hep-th]].
  %%CITATION = ARXIV:1503.07149;%%

 %\cite{Ishii:2015gia}
\bibitem{Ishii:2015gia} 
  T.~Ishii, E.~Kiritsis and C.~Rosen,
  %``Thermalization in a Holographic Confining Gauge Theory,''
  JHEP {\bf 1508}, 008 (2015)
%  doi:10.1007/JHEP08(2015)008
  [arXiv:1503.07766 [hep-th]]. 
  
 %\cite{Ryu:2015vwa}
\bibitem{Ryu:2015vwa}
  S.~Ryu, J.-F.~Paquet, C.~Shen, G.~S.~Denicol, B.~Schenke, S.~Jeon and C.~Gale,
  %``Importance of the Bulk Viscosity of QCD in Ultrarelativistic Heavy-Ion Collisions,''
  Phys.\ Rev.\ Lett.\  {\bf 115} (2015) no.13,  132301
  [arXiv:1502.01675 [nucl-th]].

%\cite{Anninos:2017hhn}
\bibitem{Anninos:2017hhn} 
  D.~Anninos and D.~M.~Hofman,
  %``Infrared Realization of dS$_2$ in AdS$_2$,''
  arXiv:1703.04622 [hep-th].
  %%CITATION = ARXIV:1703.04622;%%
  %2 citations counted in INSPIRE as of 13 Jul 2017 

   
  \end{thebibliography}
\end{document}